\documentclass[11pt]{article}

\usepackage{amsfonts}

\usepackage{amssymb}
\usepackage{latexsym}
\usepackage{graphicx}
\usepackage[english]{babel}
\usepackage[font={small}]{caption}

\usepackage{amsfonts}
\usepackage{latexsym}
\usepackage{graphicx}
\usepackage[english]{babel}
\begin{document}
\input epsf

\def\p{\partial}
\def\h{{1\over 2}}
\def\be{\begin{equation}}
\def\bea{\begin{eqnarray}}
\def\ee{\end{equation}}
\def\eea{\end{eqnarray}}
\def\d{\partial}
\def\la{\lambda}
\def\eps{\epsilon}
\def\bb{\bigskip}
\def\mm{\medskip}
\newcommand{\dm}{\begin{displaymath}}
\newcommand{\edm}{\end{displaymath}}
\renewcommand{\b}{\tilde{B}}
\newcommand{\gm}{\Gamma}
\newcommand{\ac}[2]{\ensuremath{\{ #1, #2 \}}}
\renewcommand{\ell}{l}
\newcommand{\z}{\ell}
\newcommand{\newsection}[1]{\section{#1} \setcounter{equation}{0}}
\def\bb{$\bullet$}
\def\Qbar{{\bar Q}_1}
\def\QPbar{{\bar Q}_p}

\def\q{\quad}

\def\bn{B_\circ}

\let\a=\alpha \let\b=\beta \let\g=\gamma \let\d=\delta \let\e=\epsilon
\let\c=\chi \let\th=\theta  \let\k=\kappa
\let\l=\lambda \let\m=\mu \let\n=\nu \let\x=\xi \let\r=\rho
\let\s=\sigma \let\t=\tau
\let\vp=\varphi \let\vep=\varepsilon
\let\w=\omega      \let\G=\Gamma \let\D=\Delta \let\Th=\Theta
                     \let\P=\Pi \let\S=\Sigma

\def\h{{1\over 2}}
\def\t{\tilde}
\def\r{\rightarrow}
\def\nn{\nonumber\\}
\let\bm=\bibitem
\def\Kt{{\tilde K}}
\def\b{\bigskip}

\let\p=\partial

\begin{flushright}
\end{flushright}
\vspace{20mm}
\begin{center}
{\LARGE  Nature abhors a horizon\footnote{Essay awarded third prize in the  Gravity Research Foundation 2015 essay competition.}}
\\
\vspace{18mm}
Per Kraus${}^*$ and Samir D. Mathur${}^{**}$

\vspace{8mm}

${}^*$Department of Physics and Astronomy, UCLA, Los Angeles, CA 90095-1547, USA\\
pkraus@ucla.edu

\vskip .1 in

${}^{**}$ Department of Physics, The Ohio State University, Columbus,
OH 43210, USA\\mathur.16@osu.edu\\
\vspace{4mm}
 March 31, 2015
\end{center}
\vspace{10mm}
\thispagestyle{empty}
\begin{abstract}

The information paradox can be resolved if we recognize that the wavefunctional in gravity $\Psi[g]$ should be considered on the {\it whole} of superspace, the space of possible $g$. The largeness of the Bekenstein entropy implies a vast space of gravitational solutions, which are conjectured to be  fuzzball configurations. In the WKB approximation, the wavefunctional for  a collapsing shell is oscillatory in a small region of superspace, and the classical approximation picks out this part. But the wavefunctional will be damped (`under the barrier') in the remainder of this vast superspace. We perform a simple computation to show that at the threshold of black hole formation, the barrier is lowered enough to make the latter part oscillatory; this  alters the classical evolution and avoids horizon formation.

\end{abstract}
\vskip 1.0 true in

\newpage
\setcounter{page}{1}

There are two central mysteries in black hole physics: (i) What does the enormously large entropy of a black hole signify? (ii) How can information emerge from the hole?  In this essay we will  show how  understanding the first of these puzzles also resolves the second. 

\begin{figure}[!]
\begin{center}
\includegraphics[scale=.68]{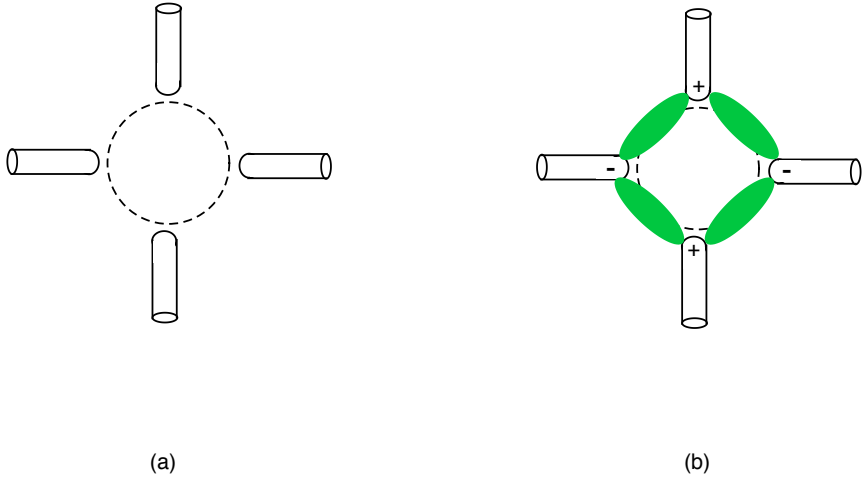}
\caption{{(a) In a `bubble of nothing', a compact direction pinches off, ending spacetime outside the bubble. (b) In a  fuzzball the structure is more complicated; the compact circle pinches off to make KK monopoles/antimonopoles (these are denoted by $+/-$ signs), and fluxes between the monopoles (shaded ovals) support the structure.  The region which would have been the interior of the horizon is absent. Different choices of fluxes and $+/-$ signs at different angular positions give the $Exp[S_{bek}]$ horizonless states of the hole.}}
\label{fone}
\end{center}
\end{figure}

Many physicists had a nagging feeling that Hawking's argument for information loss \cite{hawking} was flawed: subtle corrections could somehow encode information in the outgoing radiation. We now know, however, that the Hawking argument is stable to all small corrections; using strong subadditivity of entanglement entropy we can show that information {\it will} be lost unless physics at the horizon is altered by {\it order unity} \cite{cern}. Fortunately, string theory supports a plethora of solutions that look like black holes beyond the radius of a would-be horizon, but which are in fact horizon free. They exist because of the emergence of a new topological structure:  a compact direction in the geometry pinches off, ending the spacetime smoothly just outside the location where the horizon would have appeared (fig.\ref{fone}a) \cite{fuzzballs}. These `fuzzball' states, which have no horizon, are expected to agree in count with $Exp[S_{bek}]$, where $S_{bek}$ is the Bekenstein entropy. Their structure is reminiscent of the `bubble of nothing' \cite{witten} where a region of spacetime is removed and the pinching of a compact circle at the boundary of this region yields a geodesically complete manifold (fig.\ref{fone}b).

If black hole microstates have no horizon, then we do not have Hawking's pair creation process, and thus no information loss; the fuzzball radiates from its surface (at the Hawking rate \cite{cm1}) just like a star. But one may ask: if we start with a collapsing shell of mass $M$, then can't we just trust the classical approximation where the shell collapses through $r=2M$ to create a horizon? If such a  horizon is indeed created, then light cones inside the horizon point towards smaller $r$. Thus causality would forbid any effects that happen at $r<2M$ from altering the vacuum state at $r=2M$. It would then appear that we can never  get to a fuzzball state at the horizon, so how have we resolved the information paradox?

To see the way around this conundrum first consider a toy model, where we assume that the theory contains a very large number $N$ of massless scalar fields $\phi_i$. If
\be
N\gg \left ( {M\over m_p}\right )^2
\label{one}
\ee
then the collapsing shell will not make a black hole. Particle creation in curved space starts before the shell reaches its horizon. For each species $\phi_i$ we have $\sim 1$ quanta radiated while the shell collapses from, say, $r=4M$ to $r=2.5$ M. Each quantum has wavelength $\sim GM$ (the only length scale in the problem), so if the shell indeed reached down to $2.5 M$, then the total mass radiated in all the fields $\phi_i$ would be
\be
M_{rad}\sim {N\over GM} \gg M
\label{twop}
\ee
where in the last step we have used  (\ref{one}) and noted that $G\sim 1/m_p^2$. Thus the shell radiates away before reaching the vicinity of $r=2M$. 

In string theory we do not have a large number of massless fields, but we do have a large number of {\it massive} fuzzball states. We call these states $|F_i\rangle$; each has mass $M$, and the number of such states is  $Exp[S_{bek}]$. In \cite{tunnel} it was argued that a collapsing shell can tunnel into any given $|F_i\rangle$  at a very small rate $\sim Exp[-\alpha S_{bek}]$, where $\alpha\sim 1$. But the number of states $|F_i\rangle$ we can transition to is $Exp[S_{bek}]$. So the overall rate of transitioning to fuzzballs is
\be
\Gamma\sim N_{fuzzballs} \, Exp[-\alpha S_{bek}]\sim Exp[S_{bek}]Exp[-\alpha S_{bek}]
\label{two}
\ee
If $\alpha\le 1$, then we transition to fuzzballs quickly, and thus never make a horizon. This would remove the information paradox. 

Unfortunately, while one could estimate that $\alpha\sim 1$, there was no immediate way to show that $\alpha\le 1$. We now close this loophole by using an old result  to argue that in fact $\alpha=1$.

Let us begin by recalling Hawking's radiation rate for quanta of energy $\omega$
\be
\Gamma \sim e^{-{\omega\over T}}\sim e^{-8\pi GM \omega}
\ee
In \cite{kw} this expression was modified to include backreaction. For radiation of a spherical shell  of {\it any}  energy $\omega\le M$, we get
\be
\Gamma\sim e^{-8\pi G (M-{\omega\over 2}) \omega}
\ee
This is a special case of a general expression  valid for all holes in all dimensions \cite{kk}
\be
\Gamma\sim e^{S_{bek}(M-\omega)-S_{bek}(M)}\equiv e^{-\Delta S_{bek}}
\ee

\begin{figure}[htbp]
\begin{center}
\includegraphics[scale=.48]{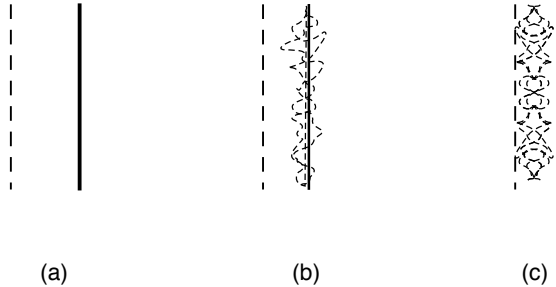}
\caption{{(a) A collapsing shell approaches its horizon (the dotted line). (b) A part of the wavefunction of the shell spreads over the space of fuzzball states. (c) At later Schwarzschild times,  the wavefunction has tunneled out of the shell state and spread completely over the space of fuzzball states; thus the shell never falls through its horizon.}}
\label{fsixdel}
\end{center}
\end{figure}

We are interested in the case where the entire mass $M$ in the shell transitions to a fuzzball state $|F_i\rangle$ of energy $\omega=M$. Thus $S_{bek}(M-\omega)=0$ in this case, and
\be
e^{-\Delta S_{bek}}\sim e^{S_{bek}(M-\omega)-S_{bek}(M)}\sim e^{-S_{bek}(M)}
\ee
Thus we see that for any one species $|F_i\rangle$ we get 
\be
\Gamma\sim e^{-S_{bek}(M)}
\ee
giving 
\be
\alpha=1
\label{ten}
\ee
Equation (\ref{ten}) is the main result of this article.  The rate at which the geometry tunnels to fuzzballs is small (as expected), but we see (eq.(\ref{two})) that this smallness is {\it exactly} cancelled by the largeness of the Bekenstein entropy, which governs how many final states we can tunnel to.   We therefore have the picture depicted in fig.\ref{fsixdel}: as the shell approaches the horizon (in the Schwarzschild frame), its state transitions to a linear combination of fuzzball states $|F_i\rangle$, at such a rate that there is no amplitude left in the shell state itself by the time the shell reaches $r=2M$. Thus the classical approximation is completely violated, and no horizon forms. This resolves the information paradox. 

To extract the moral of this story, consider again the toy model which gives (\ref{twop}). We can imagine that the collapsing shell is made of gravitons. If we consider only classical physics for all fields, then the scalars $\phi_i$ would never  be involved; the shell would pass through its horizon to create a black hole. What prevents horizon formation is the change in the quantum vacuum state of the $\phi_i$ caused by the mass $M$ in the shell; this change of vacuum leads to pair production. The largeness of $N$ (eq. (\ref{one})) makes the virtual fluctuations of the $\phi_i$  sufficiently important to modify the classical evolution of the shell completely. 

With fuzzballs, the gravitational theory has a large number $Exp[S_{bek}(M)]$ of states $|F_i\rangle$ of mass $M$. In the Minkowski vacuum, we do not have the energy to create these states on-shell. But we should not ignore the vacuum fluctuations into this very large set of  states. The infalling shell alters these vacuum fluctuations. The energy of the shell gets converted to on-shell states $|F_i\rangle$, and the classical motion of the shell is completely invalidated.

Put another way, we should always consider the wavefunctional of gravity on the {\it whole} of superspace, the space of all spatial geometries. Suppose we write the superspace wavefunctional $\Psi[g]$ of the collapsing shell  in the WKB approximation.   The wavefunctional would be oscillatory on some part of  superspace -- the classically allowed region -- but will be exponentially damped  (`under the barrier') in the rest of superspace. Classical physics focuses on the oscillatory part, but we will now show by a toy model that the `under the barrier' part of the wavefunction can become very important when there are a large number of degrees of freedom.

\begin{figure}[htbp]
\begin{center}
\includegraphics[scale=.58]{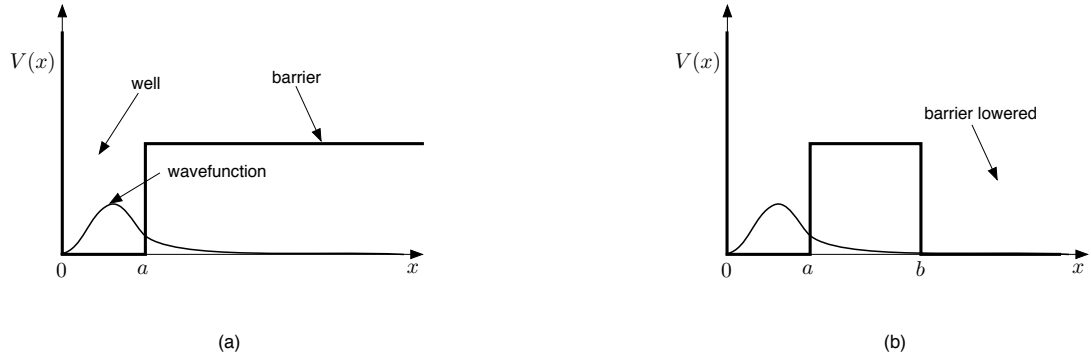}
\caption{{(a) The potential has an infinite wall at $x=0$, and a barrier starting at $x=a$. Most of the wavefunction $\psi(x)$ is located in the well $0\le x\le a$, with a small fraction under the barrier $x>a$. (b) The barrier is lowered in the region $x>b$,  so that the wavefunction in the well can slowly leak out to the region $x>b$. }}
\label{ffour}
\end{center}
\end{figure}

Consider the quantum mechanical problem of a particle in the 1-dimensional potential $V(x)$ depicted in fig.\ref{ffour}(a). Assume that 
\be
\int_{x=0}^a  dx\,  |\psi(x)|^2  = 1-\epsilon, ~~~~~\epsilon \ll 1
\ee
so that most of the norm is in the well $0\le x\le a$. Now consider the particle in  $N$ dimensions with potential
\be
V(x_1, x_2, \dots x_N)= V(x_1)+V(x_2)+\dots V(x_N)
\ee
and take
\be
 N\gg 1/\epsilon
 \ee
The wavefunction is
\be
\Psi=\psi(x_1)\psi(x_2)\dots \psi(x_N)
\ee
The  norm in the central well $0\le x_i \le a$ is now
\be
\left ( \int_{x_1=0}^a dx_1 |\psi(x_1)|^2\right )\left ( \int_{x_2=0}^a dx_2 |\psi(x_2)|^2\right )   \dots \left (\int_{x_N=0}^a dx_N |\psi(x_N)|^2\right ) =(1-\epsilon)^N \approx e^{-\epsilon N}\ll 1
\ee
so that most of the norm is {\it under the barrier}. 

To see the dynamical consequences of this fact, imagine that at $t=0$ the barrier is altered to that in fig.\ref{ffour}(b). For one copy $\psi(x)$, suppose that the wavefunction leaks out to infinity at a slow rate
\be
 \int_{x=0}^a dx |\psi(x, t)|^2  \approx (1-\epsilon) e^{-\t \epsilon t}, ~~\t \epsilon\ll 1
\ee
But for the case of $N\gg 1/\t\epsilon$ dimensions, the norm in the central well 
\be
\left ( \int_{x_1=0}^a dx_1 |\psi(x_1, t)|^2\right )\left ( \int_{x_2=0}^a dx_2 |\psi(x_2, t)|^2\right ) \dots  \left ( \int_{x_N=0}^a dx_N |\psi(x_N, t)|^2\right ) \approx (1-\epsilon)^N e^{-N\t \epsilon t}
\ee 
decays almost instantly, in a time 
\be
\Delta t \sim 1/(N\t\epsilon)\ll 1
\ee

Let us summarize. It was known that black holes are characterized by a remarkably large entropy $S_{bek}$, but we did not know what this largeness signified for the dynamics of the hole. With the advent of the fuzzball construction it is now conjectured that $Exp[S_{bek}]$ counts explicitly constructable states -- fuzzballs -- which are normal quantum gravitational states without horizon. Under normal conditions (i.e. when we are not close to making a black hole)  these states lie in the regime of superspace where the wavefunction is `under the barrier', so we do not directly see these states. But as a collapsing shell approaches its horizon, the wavefunctional tunnels into the ensemble of these states at rate that is {\it not} suppressed; the smallness of tunneling rate to each fuzzball is exactly cancelled by the largeness of the number of fuzzballs. The norm of the wavefunction captured by these fuzzball states thus grows to unity, just as in the toy example of a particle in an $N$ dimensional potential well with $N$ large. This tunneling alters the classical dynamics of a collapsing shell completely and resolves the information paradox, since the fuzzballs that result have no horizon.
 
It is now interesting to ask how this altered dynamics should be described; in particular we have a tantalizing conjecture of fuzzball complementarity \cite{comp} which says that collective oscillations of the fuzzball surface approximately mimic, in a {\it dual} description, the dynamics of the black hole interior. This approximation allows information to escape in  low energy $E\sim T$ quanta of Hawing radiation, while dynamics of $E\gg T$ quanta mimics  classical physics in the dual picture. 
   
\section*{Acknowledgements}

We would like to thank Emil Martinec for many helpful discussions.  SDM is supported in part by DOE grant de-sc0011726.

\end{document}